\documentclass[12pt,twocolumn]{article}
\usepackage{graphicx}
\usepackage{hyperref}
\usepackage{pstricks,amssymb}
\usepackage{fancyhdr}
\usepackage[raggedright]{titlesec}
\begin{document}
\title{DARK MATTER AND NEUTRINOS\\
\vspace{.3cm}
\large{Gazal Sharma${^1}$, Anu${^2}$ and B. C. Chauhan${^3}$}\\
\vspace{0.3cm}
\normalsize{
Department of Physics \& Astronomical Science\\
School of Physical \& Material Sciences\\
Central University of Himachal Pradesh (CUHP)\\
Dharamshala, Kangra (HP) INDIA-176215.\\
$^1{}${\small gazzal.sharma555@gmail.com}
$^{2}${\small 3anoman7@gmail.com}\\
$^{3}${\small chauhan@associate.iucaa.in}\\
\vspace{0.3cm}
(Submitted 03-08-2015)}\\
\rule[0.1cm]{16cm}{0.02cm}\\
\textbf{Abstract}\\
\flushleft\normalsize{The Keplerian distribution of velocities is not observed 
in the rotation of large 
scale structures, such as found in the rotation of spiral 
galaxies. The deviation from Keplerian distribution 
provides compelling evidence of the presence of 
non-luminous matter i.e. called dark matter. There are several
astrophysical motivations for investigating the dark matter 
in and around the galaxy as halo. In this work we address 
various theoretical and experimental indications pointing 
towards the existence of this unknown form of matter. 
Amongst its constituents 
neutrino is one of the most prospective candidates.
We know the neutrinos oscillate and have tiny masses, 
but there are also signatures for existence of heavy and 
light sterile neutrinos and possibility of their mixing. 
Altogether, the role of neutrinos is of great interests 
in cosmology and understanding dark matter.}\\
\rule[0.1cm]{16cm}{0.02cm}
}
\date{}
\maketitle
\thispagestyle{fancy}
\lhead{\textbf{Physics Education}}
\chead{\thepage}
\rhead{\bf{dateline}}
\lfoot{Volume xx, Number y Article Number : n.}
\cfoot{}
\rfoot{www.physedu.in}
\renewcommand{\headrulewidth}{0.4pt}
\renewcommand{\footrulewidth}{0.4pt}

\section{Introduction}
As a human being the biggest surprise for us was, that the Universe in which we live in is mostly dark. The NASA's Plank Mission revealed in 2013 that our Universe contains $68.3\%$ of dark energy, $ 26.8\%$ of dark matter, and only $4.9\%$ of the Universe is known matter which includes all the stars, planetary systems, galaxies, and interstellar gas etc.. This raises a number of questions in our minds; e.g. how much and how well we know about our Universe? What are dark matter and dark energy? What are they made up of? The very first suggestion of dark matter in our galaxy was made by Kapteyn and Jeans in 1922 and then by Lindblad in 1926. They proposed the existence of dark matter while observing the motions of nearby stars at right angle to the plane of our Milky way galaxy. Oort in 1932 claimed that there exists substantial amount of dark matter near the sun by observing the vertical motions of stars. However, in 1991, Kuijken and Gilmore argued that there were no significant evidence for dark matter with in the galactic disk near the sun. 

Sinclair Smith and Fritz Zwicky in 1933, studied the large clusters of galaxies and found that  galaxies were on average moving too fast for the cluster to be held together only by the mass of the visible matter. They concluded that in rich clusters of galaxies, a large portion of the matter is not visible i.e. the dark matter. The idea of dark matter in galactic halo was given by Freeman in 1970, while 
studying the rotation curve for NGC 300 and M33 by using the 21cm-Line of neutral hydrogen did not show the expected Keplerian decline beyond the optical radii. Then in 1979, Vera Rubin proposed that normal spiral galaxies contain substantial amount of dark matter present at great distances from the central regions. An influential model was proposed by Caldwell and Ostriker in 1981 for the density of core-halo type model of dark matter. The halo model is valid till now but the exact distribution of  dark matter is still a mystery. 

The next question to be addressed is about the constituents of dark matter. 
One of the biggest discoveries made by Hubble Space Telescope (HST) of NASA was the confirmation of invisible matter in the Universe. A 3D map of dark matter was derived from largest survey of the Universe made by the HST, the Cosmic Evolution Survey (COSMOS). The COSMOS survey covers a sufficiently wide area of sky - nine times the area of the full Moon (1.6 square degrees) - for the large-scale filamentary structure of an invisible form of matter that makes up most of the mass of Universe i.e. dark matter to be clearly evident \cite{HST}. 
The theory of Big-Bang nucleosynthesis (BBN), i.e. formation of light nuclei just after Big-Bang, as well as experimental evidences from anisotropies in Cosmic Microwave Background Radiation (CMBR) observed by NASA's Wilkinson Microwave Anisotropy Probe (WMAP) indicate that most of the dark matter stuff is non-baryonic (which is not made up of regular matter).

Many experiments has been performed in search of dark matter candidates.
Neutrinos, which are electrically neutral and tiny particles, seem potential candidates for dark matter, as they are long-lived and almost non-interacting with other particles. However, the three known types of neutrinos, called active neutrinos, are not massive enough to account for all of the dark matter of Universe. So, theorists proposed another type of neutrinos that would not interact at all with the regular matter, but are massive. 
If the sterile neutrino is heavy enough about $\sim 10~keV$, it could account for the substantial amount of dark matter. The present article aims to introduce reader about the dark matter, its evidences, possible constituents and the potential candidature of neutrinos in the composition of dark matter.

\section{Dark Matter}
As discussed above the dark matter is the matter, which does not interact with light 
at all or may interact very poorly that it remains dark and unseen. As such, a question arises in ones mind; how one can detect something which does not interact with light. The answer may be 'gravity'; such that there are many astrophysical motivations for the detection of dark matter. There have been obtained a number of observational evidences for the existence of dark matter because of its gravitational effects, like galactic rotation curves of galaxies measured by Vera Rubin, confinement of hot gas in the galaxies, measurement on the basis of gravitational lensing \cite{c}, etc...

\subsection{Flattened Orbital Velocity Curves}
Before describing observations let us see how celestial 
objects respond to the  gravitational force acting on them 
and how that response can reveal the large scale 
distribution of matter. For the planets in orbit around 
the sun which embodies essentially all the mass of the 
solar system, the decrease in gravitational attraction 
with distance is given by Newton law of gravitation. 
It has been found that 
orbital velocities of planets decreases with distance from 
the centre of the sun. In spiral galaxy the gas, dust and 
stars in the disk of the galaxy all orbit around a common 
centre. Like planets in solar system, the gas and stars 
move in response to the combined gravitational attraction  
of all other mass. 
If the galaxy is visualized as a spheroid, we can 
calculate the gravitational attraction due to mass 
\textit{$M_{r}$}  lying between the centre and an object 
of mass \textit{m}  in an equatorial orbit at a distance 
\textit{r} from the centre. If the galaxy is neither 
contracting nor expanding then the gravitational force is 
exactly equal to the centrifugal force on the mass at 
distance \textit{r} is given by the equation
\begin{equation}  
GmM_{r}/r^{2} = mv_{r}^{2}/r,                
\end{equation}

where $v_{r}$ is the orbital velocity. When the equation 
is solved for $v_{r}$ , the value of \textit{m} drops out 
and the velocity of a body at a distance \textit{r} from 
the centre is determined only by the mass $M_{r}$ inward 
from its position. In the solar system, virtually all the 
mass is concentrated near the centre and the orbital velocity 
decrease as $1/ \surd{r}$, that is called Keplerian decline.

\begin{figure}[hbtp]
\caption{\textit{{Variation of orbital velocity with radius}} 
\cite{ad}}
\centering
\includegraphics[scale=0.9]{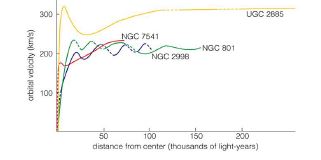}
\end{figure}

In a galaxy the brightness is strongly peaked near the 
centre and falls off rapidly with distance. As per the 
distribution of luminosity,  it was expected that stars at 
increasing distance from the centre would have decreasing 
Keplerian orbital velocities. When the orbital velocity of 
different stars present at different distances in a galaxy 
was studied by Vera Rubin, the unexpected results came out. 
This observation has been made for different spiral 
galaxies like Sa, Sb, Sc etc... Although each  galaxy 
exhibits distinctive feature in its rotational pattern, 
the systematic trends that emerge are impressive. With 
increasing luminosity galaxies are bigger, orbital 
velocities are higher and the velocity gradient across the 
nuclear bulge is steeper. Moreover, each type of galaxy 
displays characteristic rotational properties. 

Therefore we can draw conclusion from our observation that 
all the rotation curves are either flat or rising out to 
the visible limits of the galaxy. There are no extensive 
regions where the velocities fall off with distance from 
the centre, as would be predicted if mass were centrally 
concentrated. The conclusion is inescapable- mass unlike 
luminosity is not concentrated near the centre of spiral 
galaxies. Thus the light distribution in a galaxy is not 
at all a guide to mass distribution. Instead the mass 
inside any given radial distance is increasing linearly 
with distance and contrary to what one might expect, is 
not converging to a limiting mass at the edge of the 
visible disk. The linear increase in mass with radius 
indicates that each successive shell of matter in the 
galaxy must contain just as much mass as every other shell 
of the same thickness. Since the volume $V$ of each 
successive shell increases as the square of the radius, 
the density $\rho$ of matter in successive shells must 
decrease in order to keep $\rho V$ constant \cite{a}.

The widely accepted idea about the dark matter is that 
each spiral galaxy is embedded into a halo of dark matter. 
The gravitational attraction of the unseen mass keeps the 
orbital velocities high at larger distance from galactic 
centre. Till now we are not able to find the exact 
distribution of dark matter but we can say that it is 
strongly clumped around the galaxies. The density of dark 
halo decreases with distance from galactic centre as given 
by Caldwell and Ostriker
\begin{equation}
\rho_{d} = \frac{\rho_{0}}{1+\frac{r^2}{a^2}}.
\end{equation}
They found a fit for the data with $\rho_{0}=1.37\times 
10^{-2} M_{\odot} pc^{-3},$ and $a=7.8 kpc$.

If we consider a different distribution for dark matter 
in which we put all the unseen matter in a disk, the disk 
will quickly become unstable. Therefore P. Ostriker and 
Peeble suggest that the halos are important for stabilizing the 
disk. Additional evidence on the high rotational velocity 
was provided  by the 21-centimetre radio waves emitted by 
the neutral hydrogen in the galactic disk. From the above 
discussion we 
can draw a conclusion that the density of dark matter halo 
surrounding the visible matter decreases slowly outwards.

\subsection{Gravitational lensing}
According to Einstein's theory of general relativity large 
objects with their immense masses can distort space-time 
therefore large massive objects such as galaxy clusters 
bend light from distant sources, creating distorted images 
that we can see here on earth. This is called 
gravitational lensing. This technique is especially useful 
for detecting dark matter. Since dark matter doesn't 
interact with light, it can't be seen directly. However, 
since dark matter is very massive, it can be detected 
indirectly by the distorted images it creates of normal 
matter through gravitational lensing. By measuring the 
angle of bending, the mass of the gravitational lens can 
be calculated- greater the bend, more massive the lens is. 
Therefore the angle of deflection is given by \cite{i}
\begin{equation}
\alpha = \frac{4GM}{c^{2}b},
\end{equation}

where $\textit{b}$ is the impact parameter. Using this 
method, astronomers have confirmed that the galactic 
clusters indeed have high masses exceeding those measured 
by the luminous matter. There have been several positive 
reports on the observation of such micro-lensing, 
even though typically only one in a million stars examined 
is expected to show such an effect. The bending of light 
by a massive object, a general relativity effect has been verified 
to extreme accuracy (better than $1\%$) by studying radar 
echoes from the planets when they are in conjunction. 
Experiments like the Large Synoptic Survey Telescope (LSST),
under construction in Chile, aim to take advantage of 
gravitational lensing to map the 
dark matter in the Universe and provide clues to its nature. 
MOA (Micro-lensing Observations in Astrophysics) is a 
Japan/NZ collaboration that makes observations on dark 
matter, extra-solar planets and stellar atmospheres using 
the gravitational micro-lensing technique at the Mt John 
Observatory in New Zealand. HST of NASA recently produces 
several images of gravitational lensed objects. 
Therefore finding enough gravitational lenses to constrain 
the properties of dark matter structures requires a 
powerful telescope with a huge field of view like LSST.  

\subsection{Fluctuations in CMBR}
The Cosmic Microwave Background (CMB) is the earliest 
photograph of our Universe. The patterns that we see in 
observations of the CMB were set up by competition between 
two forces acting on matter; the force of gravity causing 
matter to fall inward and an outward pressure exerted by 
photons (or particles of light). This competition caused 
the photons and matter to oscillate into-and-out-of dense 
regions. If the Universe consisted partially of dark 
matter in addition to normal matter, that pattern would be 
affected dramatically. The existence of dark matter leaves 
a characteristic imprint on CMB observations, as it clumps 
into dense regions and contributes to the gravitational 
collapse of matter, but is unaffected by the pressure from 
photons. We can predict these oscillations in the CMB with 
and without dark matter, which often present in the form 
of a power spectrum. 

The power spectrum of the CMB shows us the strength of 
photon-matter oscillations at different parts of sky. 
The Far-Infrared Absolute Spectrophotometer (FIRAS) 
instrument has measured the spectrum of the cosmic 
background radiation, making it the most precisely 
measured black body spectrum in nature. The Cosmic 
Background Explorer (COBE) was launched in 1989 in search 
of temperature anisotropies; frequency power spectrum; 
solar system and galactic dust foregrounds. The WMAP in 2010 
was the first instrument to measure the CMB power spectrum 
through the 
first peak of oscillations, and showed that the existence 
of dark matter is favoured. Comparison of such calculations 
with the observations of CMB Radiation by Plank mission team in 
$2013$ have shown that the total mass energy of the known 
Universe contains $4.9\%$ ordinary matter,  $26.8\%$ dark 
matter and $68.3\%$ dark energy. Thus dark Universe 
constitutes  $95.1\%$ of the total matter energy content 
of the Universe \cite{b}.

\subsection{X- Ray Studies}
The observational evidences from X- ray studies also 
supports the existence of dark matter. The basic technique 
is to estimate the temperature and density of the gas from 
the energy and flux of the X-rays using X-ray telescopes, 
which would further enable the mass of the galactic 
cluster to be derived. The measurements of hot gas 
pressure in galactic clusters by X-ray telescopes, such as 
CHANDRA X-ray observatory by NASA, have shown that  the 
amount of superheated gas is not enough to account for the 
discrepancies in mass and that the visible matter 
approximately constitutes only $12-15\%$ of the mass of 
the cluster. Otherwise, there won't be sufficient gravity 
in the cluster to prevent the hot gas from 
escaping \cite{b}.

Recently in 2014, data came from the European Space 
Agency's (ESA's) XMM-Newton spacecraft, which was 
analysed by an international team of researchers. After 
scouring through thousands of signals, they spotted a 
weird spike in X-ray emissions coming from two different 
spots in the Universe: the Andromeda galaxy and the 
Perseus galaxy cluster. The signal doesn't correspond to 
any known particle or atom, and is unlikely to be the 
result of a measurement or instrument error hence it could 
have been produced by a dark matter particle. The 
signal's distribution within the galaxy corresponds 
exactly to what we expects with dark matter i.e. 
concentrated and intense in the centre of objects and 
weaker and diffuse on the edges. Scientists believe that
there is a possibility that it could come from dark matter candidate i.e. possibly the hypothetical heavy sterile neutrinos; as it is believed the decay of these particles could produce X-rays \cite{n}.

\section{Baryonic \& Non-Baryonic Dark Matter}
On the basis of observed orbital velocity curves, 
and other evidences we can 
say that dark matter exists. Baryonic dark matter is 
non-luminous matter in which most of the mass is 
attributed to baryons, most probably neutrons and protons. 
Candidates for baryonic dark matter include non-luminous 
gas, Massive Astrophysical Compact Halo Objects (MACHOs). 
These MACHOs may include condensed objects such as black 
holes, neutron stars, white dwarf, very faint stars, or 
non-luminous objects like planets and brown dwarfs.
Baryonic dark matter cannot be detected by its emitted 
radiation because these objects have very low luminosity, 
but the presence of these objects can be inferred from 
their gravitational effects on visible matter \cite{b}. 

The nucleosynthesis of the elements and observations of 
the Cosmic Microwave Background Radiations (CMBR)  puts 
constraints on the density of baryonic matter. No more 
than $15\%$ of the matter in the Universe can be baryonic 
but most of dark matter stuff is non-baryonic. 
Non-baryonic dark matter (NBDM) is non-luminous matter 
made from non-baryonic stuff (other than protons, 
neutrons etc.). Recent measurements of the matter density 
$\Omega_{m}^{0}$ and the energy density 
$\Omega_{\Lambda}^{0}$ comes from three types of 
observations: 1) supernova measurements of the recent 
expansion history of the Universe; 2) cosmic microwave 
background measurements of the degree of spatial flatness, 
and 3) measurements of the amount of matter in galaxy 
structures obtained through big galaxy redshift surveys 
agree with each other in a region around the best 
current values of the matter and energy densities 
$\Omega_{m}^{0}$ $\simeq 0.27 $ and $\Omega_{\Lambda}^{0} 
\simeq 0.73$. Where $\Omega$ is the energy density of 
Universe defined by
\begin{equation}
\Omega = \frac{\rho}{\rho_{c}},
\end{equation}

where $\rho_{c}$ is the critical density (average density 
of Universe to halt its expansion) of the Universe and 
$\Omega^{0}$ represents  present energy density of 
Universe. Measurements of the baryon density in the 
Universe using CMBR spectrum and primordial 
nucleosynthesis (i.e. BBN) constrain the 
baryon density $\Omega_{b}^{0}$ to a value less than 0.05. 
The difference $ \Omega_{m}^{0} - \Omega_{b}^{0} \simeq 
0.22 $ must be in form of non-baryonic dark matter \cite{ae}. 
The value of total matter density
\begin{equation}
\Omega_{m}^{0}h^{2} = 0.135^{+0.008}_{-0.009},
\end{equation}
out of which the baryonic matter is
\begin{equation}
\Omega_{b}^{0}h^{2} = 0.0224^{+0.0009}_{-0.0009},
\end{equation}

in the form of neutrinos
\begin{equation}
\Omega_{\nu}^{0} h^{2} < 0.0076, 
\end{equation}

and the matter in the form of Cold Dark Matter (CDM) is
\begin{equation}\nonumber
\Omega_{CDM}^{0} h^{2} = 0.113 ^{+0.008}_{-0.009}. 
\end{equation}

The results of BBN that 
tell that $ \Omega_{B} \sim  0.01$  and therefore if  $ 
\Omega $ total is truly unity, then the bulk of the mass 
of the Universe must be in the form of some sort of 
non-baryonic matter. From baryon to photon ratio i.e. 
$\eta = \eta_{B}/\eta_{\gamma}$,  one can find the range 
for $\eta$ as given by \cite{g}
\begin{eqnarray}
4.7\times 10^{-10} \le \eta \le 6.5 \times 10^{-10}.
\end{eqnarray}
We can find relative baryon density $\Omega_{B}$ as

\begin{eqnarray}
0.017 \le \Omega^0_{B}h^{2} \le 0.024.
\end{eqnarray}

This shows that Universe is not closed by baryonic matter 
and this gives the indication of existence of dark matter. 
From the analysis of the existing data follows that
\begin{equation}
\Omega_{DM}\simeq 0.20.
\end{equation}

The non-baryonic dark matter is classified in terms of 
the mass of the particle that is assumed to make it up, 
and the typical velocity dispersion of those particles 
(since more massive particles move more slowly). There are 
three prominent hypotheses on non-baryonic dark matter, 
called Hot Dark Matter (HDM), Warm Dark Matter (WDM), and 
Cold Dark Matter (CDM); some combination of these is also 
possible. CDM is composed of substantially massive 
particles $(\sim GeV)$ expected to be moving with 
non-relativistic speeds. The leading candidates for CDM 
called WIMPs (Weakly Interacting Massive Particles). WIMPs 
could include large number of exotic particles such as 
neutralinos, axions, photinos etc. These particles forms 
dark matter, because they have too much mass to move at 
high speeds and that they are the best candidates for dark 
matter. As WIMPs can interact through gravitational and weak forces only,
they are extremely difficult to detect. There are several experiments setup 
for detection of WIMPs such as SuperCDMS,  NASA's Fermi 
Gamma-Ray Space telescope, Large Hydron Collider (LHC) 
at Geneva etc... Experimental efforts to detect WIMPs include the 
search for products of WIMP annihilation, including gamma 
rays, neutrinos and cosmic rays in nearby galaxies and 
galaxy clusters; direct detection experiments designed to 
measure the collision of WIMPs with nuclei in the 
laboratory, as well as attempts to directly produce WIMPs 
in colliders, such as the LHC. However all the efforts in 
this direction has been fruitless so far.

The HDM consists of particles to be moving nearly at 
the speed of light, when the pre-galactic clumps began to 
form. HDM includes massive $(\sim eV)$ neutrinos. The 
neutrinos are the only hot dark matter candidate as they are 
light enough to  move with the speed of light. The 
Universe is full of neutrinos left over from just after 
the Big-Bang, when matter and anti-matter were formed. 
There are huge amount of neutrinos, that if they have just 
a tiny mass, then they can significantly account for the 
dark matter. The dark matter that has properties 
intermediate between those of hot dark matter and cold 
dark matter named as Warm Dark Matter (WDM). WDM is composed 
of sub-relativistic particles having masses $(\sim keV)$ 
causing structure formation to occur bottom-up (micro to macro scale)from above 
their free-streaming scale, and top-down (macro to micro scale) below their free
streaming scale. The most common WDM candidates are considered to be sterile neutrinos and 
gravitinos. The WIMPs when produced non-thermally could be candidates for WDM \cite{p}.

The most widely discussed models for non-baryonic dark 
matter is based on the CDM hypothesis. CDM leads to bottom-up 
formation of structure in the Universe i.e. small scale structures led to the 
formation of large scale structures. On the other hand, the HDM results in top-down formation scenario i.e. first super-cluster formed and then galaxies and then the formation of small structure takes place.
However, WDM has intermediate role in large scale structure formation.

\section{Neutrino Dark Matter}
Neutrinos are most abundant particles in the Universe. 
They are electrically neutral and have tiny mass. Out 
of four interactions in nature neutrinos interact only via 
the weak interaction and feebly via gravitational force. They rarely interact with any 
material, which makes experimental detection of these 
particles extremely challenging. There are three types 
of neutrinos so far detected, which are denoted as
electronic ($\nu_{e}$), muonic ($\nu_{\mu}$), and tauonic 
($\nu_{\tau}$) flavour eigenstates. In fact, in the Standard Model 
of particle physics, neutrinos are massless. However, in 
the late 90's and beginning of $21^{st}$ century, 
physicists observed neutrino oscillation, 
a quantum mechanical effect which would not occur 
unless neutrinos have mass. The theory of neutrino oscillation 
describes the flavor eigenstates as the mixing or 
linear superposition of mass eigenstates 
$\nu_{1}$, $\nu_{2}$, $\nu_{3}$. 
For two flavour case the mixing is shown as 
\begin{equation}
\left(
\begin{array}{c}
\nu_{e}\\
\nu_{\mu}
\end{array}
\right)
=
\left(
\begin{array}{cc}
\cos\theta & \sin\theta\\
-sin\theta & \cos\theta
\end{array}
\right)
\left(
\begin{array}{c}
\nu_{1}\\
\nu_{2}
\end{array}
\right),
\end{equation}

where $\theta$ is a mixing angle. From the observation of the neutrino oscillations phenomenon, it is confirmed that neutrinos have mass. The nature of neutrinos 
is not yet understood i.e. whether they are Dirac or Majorana particles. In case of Dirac nature neutrino and antineutrino are different, while in the Majorana nature they are the same particle. Despite the tininess, the neutrino mass has far-reaching implications in astrophysics and cosmology. 

Neutrinos are considered to be the constituent of dark matter via thermal mechanism. As discussed above the hot dark matter is the matter that was relativistic until just before the epoch of galaxy formation, neutrinos of very low mass are strongest candidates for hot dark matter. It is believed that neutrinos were in thermal equilibrium with the hot plasma which filled the early Universe. As the Universe expanded and cooled, the rates of weak interaction processes decreases and neutrino decoupled when these rates became smaller than the Hubble expansion rate. Since for the three known light neutrinos with masses smaller than $1eV$, the decoupling occurred when they were relativistic called hot relics. As their interaction of cross section with matter is very small therefore, the direct detection of these relativistic neutrinos is an extremely difficult task. In 
early Universe, when $1MeV \leq T_{\gamma} \leq 100 MeV$, 
neutrinos were kept in equilibrium with primordial plasma 
by the weak interactions. The reactions of neutrinos with 
nucleons were negligible, because the number density of 
the non-relativistic nucleons was much smaller than the 
density of relativistic electrons and positrons. The 
interaction rate for each neutrino is given by\cite{h}
\begin{equation}
\Gamma \hspace{0.1cm} = n<\sigma v>,
\end{equation}

where $\textit{n}$ is the number density of target 
particles, $\sigma$ is the cross-section and $v$ is the 
neutrino velocity. The bracket denote the thermal 
averaging. For weak interaction processes
\begin{equation}
<\sigma v> \hspace{0.1cm} = G_{F}^{2}T_{\gamma}^{2},
\end{equation}

where the temperature ($T_{\gamma}$) gives the order of 
magnitude of the energies of the relativistic particles 
participating in the reactions. As the number density 
of relativistic particles is given as $n\sim T_{\gamma}^{3}$, the 
interaction rate for each neutrino became
\begin{equation}
\Gamma \sim G_{F}^{2}T_{\gamma}^{5}.
\end{equation}

So we can say that interaction rate decreases rapidly with 
the decrease of the temperature due to expansion of the 
Universe and we obtain the decoupling temperature for 
neutrinos $T_{\gamma}^{\nu} \sim 1 MeV$. 

If the active neutrinos have a non-zero mass, as indicated by 
several neutrino oscillation experiments, the sterile neutrinos will take part in the 
neutrino oscillations. The sterile neutrinos are 'sterile' as they practically inactive, and they don't interact via any other interactions with active neutrino except by mixing \cite{l0}. This allows a possibility for a radiative decay 
under emission of an X-ray photon with energy of half the sterile neutrino mass. However, it needs much more confirmation before one accepts this as the explanation.

The sterile neutrino was originally proposed as a dark 
matter candidate by Dodelson and Widrow in $1993$ to solve 
the discrepancies between the CDM predicted structure 
formation and observations \cite{g}. Since neutrinos 
were relativistic at the time of decoupling, the number density of 
relic neutrinos is given by the relativistic expression 
independent from the values of their masses. In other words, 
light neutrinos are hot relics and contribute to the hot 
dark matter in the Universe. Sterile neutrinos have been invoked to generate masses for light neutrinos; as such the mix with light neutrinos and hence can be produced via oscillations \cite{g}. With this mechanism, their relic density is estimated to be
\begin{equation}
\Omega_{N}\approx \hspace{0.1cm} \left(\frac{\sin^2\theta}{3\times10^{-9}}\right) \left(\frac{M_N}{3 keV}\right) ^{1.8}.
\end{equation}

Here, $\theta$ is the mixing angle between the sterile neutrinos $N$ with mass $M_N$ and the active neutrinos. It has been seen that a viable sterile neutrino to be the dark matter candidate requires a mass of $keV$ and a very small mixing angle. It is a WDM  
candidate and its interactions are dominated by gravity, as preferred by the 
structure formation \cite{l}.

Neutrinos with masses much smaller 
than the effective neutrino temperature are still 
relativistic and have negligible contribution toward the 
energy density of the Universe. Despite the second most abundant particles after the photons, neutrinos fail to accommodate the observed abundance of dark matter. The relic density of light neutrinos is fixed as  \cite{h}
 \begin{equation}
\Omega_{\nu}^{0}h^{2} \hspace{0.1cm} = \frac{ 
\sum_{i}m_{i}}{94.14 eV}.
\end{equation}

Thus, the neutrino energy density is proportional to the 
sum of neutrino masses. This value is relevant for the 
present energy balance if there are neutrinos with masses 
of the order of $1 eV$ or more. Before the neutrino 
decoupling around $T_{\gamma} \simeq 1 MeV$, the weak 
processes were  in equilibrium. 

$\Omega >1$, implies a closed Universe, which means that 
at some time the gravitation attraction will stop the 
expansion and Universe will collapse again. An $\Omega < 
1$, means a Universe which expands forever. However 
$\Omega= 1$ means a flat Universe. At present time,  
$\Omega $ is changing on time scale of seconds. Since our 
existence is not compatible with the Universe which is 
either closed or continuously expanding, the only long 
term value that $\Omega$ is close to unity. Although the 
detailed physical mechanism for driving the expansion is 
not well determined and differs in different grand unified 
theories. 

The phenomenon of sudden and fast expansion of Universe caused by a scalar field present in the nascent stage is known as 'inflation'. Inflation provides a possible mechanism to set the initial conditions. 
From the inflation paradigm, it is the argument that the only long lived natural value for  $\Omega$ is unity  and that inflation provided the early Universe with the mechanism to achieve 
that value and thereby solve some of the main problems of standard model of cosmology; e.g. the flatness and smoothness problems. 

The WMAP-7 data provides a quite stringent constraint on the sum of neutrino masses of
$ \sum m_{\nu} < 1.3~eV$ at 95\% c.l. \cite{WMAP7}, which is more constrained than $\approx 2.1 eV$, that is the first releases \cite{Fuku}.
However, the most recent and sophisticated analysis of Lyman-a data gives an upper bound of 0.9 eV for the sum of neutrino masses.
In summary, at present the bound on the sum of neutrino masses can be in the range between 0.3 and
more than 2 eV, depending on the data and parameters used. 
The bound can be relaxed somewhat when more parameters, such as sterile neutrinos $(\nu_{s})$ are included. In the most conservative case the bound is above 2.5 eV if only CMB data is used.
When CMB data is combined with LSS data in the linear or almost linear regime, combined with a prior
on the Hubble parameter the upper bound is robustly below 1 eV. This is true even for extended models.
Here it should perhaps also be noted that the bound on neutrino mass from cosmic structure formation
applies to any other, hypothetical particle species which decouples while still relativistic. This could for example be low mass sterile neutrinos. It could also be relatively high
mass axions which decouple after the QCD phase transition.

Neutrinos have a kinematical advantages over the dark matter candidates is that they cluster on large scales, where the dark matter is needed to hold the large clusters of galaxies. 
In HDM, since they decoupled at a temperature of the order 
of $1 MeV$ when they were relativistic and formed 
relativistic HDM gas. The HDM perturbations within the 
horizon are erased by free streaming (i.e. the random particle 
velocities close to the velocity of light disperse all HDM 
over-densities). Free streaming ceases when the HDM  gas 
becomes non-relativistic at some red-shift $Z_{nr}$. Thus, only the HDM perturbations with 
wavelength larger than the horizon distance at $Z_{nr}$ survive and can take part in 
the generation of structure in the Universe. Since the 
horizon distance at $Z_{nr}$ is typically much larger than 
the volume corresponding to the galactic size masses, so 
in a Universe dominated by HDM, the formation of 
structures must proceed according to top-down mechanism. However 
the observed statistical properties favours bottom-up 
mechanism i.e. small structures leading to the 
formation of large scale structures. Hence the HDM contribute to the formation of small scale structures while CDM is responsible for binding of large scale structures \cite{h}.

The standard model does not predict any masses for the active
neutrinos, but as stated above the masses are required by the experimentally
verified neutrino oscillations. A simple way to incorporate 
the neutrino masses is to extend the model with the right-handed 
neutrinos just as done for the other elementary particles of SM. It is possible to add an 
arbitrary number of sterile neutrinos, but at least three sterile neutrinos are
needed to explain the neutrino oscillations, the baryon asymmetry,
and the dark matter \cite{h0}. The successful 'three sterile neutrinos' extension of 
the standard model is called the (Neutrino Minimal Standard Model)($\nu$MSM). 
It is re-normalisable and in agreement with most particle physics experiments \cite{h1}. 
The Big-Bang production of $^{4} He$ increases with $\eta$. 
Thus upper limit to $^{4}He$ abundance and a lower limit to 
baryon density lead to an upper limit to  number of neutrino 
species $N_{\nu}$ (i.e. so called BBN bound). The lower limit to baryon density is 
based on the Big-Bang production of deuterium $^{2}H$, 
which rises rapidly with decreasing baryon density. Since 
all the neutrons end up in forming $^{4}He$, which is the 
most tightly bound stable light nucleus, the mass fraction 
of $^{4}He$ is denoted as $Y_{p}$, and is given by \cite{h}
\begin{equation}
Y_{p} \simeq \left(\frac{2n_{n}}{n_{n}+n_{p}}\right) 
\simeq 0.25.
\end{equation}

As per recent estimates of $Y_{p}$ with conservative 
assumptions - for $^{3}He$ chemical evolution and $Y_{p}= 0.252$, 
less than four neutrino species are possible; however, for extreme 
assumption- no limit to primeval deuterium- less than five 
neutrino species are allowed which implies there exist 
fourth neutrino flavor that is sterile neutrino. 
In summary, there are healthy signatures for additional degrees of freedom $N_{\nu}>3$ i.e. the species of sterile neutrinos from various studies, which are given below \cite{Holanda}:
$2.98< N_{\nu} <4.48$ [BBN](68\% CL); 
$3.03< N_{\nu} <7.59$ [WMAP5+SDSS-DR7+Ho ] (95\% C.L.);
$3.46< N_{\nu} <5.20$ [WMAP7+BAO+New Ho](68\% CL);
$4.0< N_{\nu} <6.6$ [WMAP7+ACT data](68\% C.L.);
$2.22< N_{\nu} <9.66$ [WMAP3](68\% CL).

Using recombination-era observables including the CMB, the shift parameter RCMB and the sound horizon from Baryon Acoustic Oscillations (BAO) severely constrain the sterile neutrino  $\sin2\theta <0.026(m_s/eV)^-2.$\cite{Olga}. 
Recent bounds on the mixing between the active and the sterile neutrinos have been derived from the combination of neutrino oscillation data and direct experimental searches for sterile neutrinos.\cite{Oleg}
Electron neutrino-sterile neutrino mixing bound \cite{Smirnov} from joint fits of solar, KamLAND, Daya-Bay and Reno experiments is $\sin2\theta_{es} <0.2.$ and the analysis of cosmological data in terms of $\Lambda CDM$ constrains the mass square difference with one sterile family $\Delta m^2_{41}< 0.25 eV^2$.  

\section{Conclusions}
Ever-since the dawn of civilisation man has been fascinated
by the stars, planets and other heavenly objects, 
wondering what essentially the magnificent Universe was 
made up of. We learnt that our Universe is almost completely
dark. To understand this mystery was the main thrust to 
know more about the invisible Universe.
The story of dark matter
began nearly a century ago, when Kapteyn and Jeans 
propounded of existing a such kind of weird matter. 
Later, Smith and Zwicky discovered 
that in some large clusters of galaxies the individual 
members are moving so rapidly that their mutual gravitational
attraction 
is insufficient to keep the clusters from flying apart. 
Either such clusters should be dissolving or there must be 
enough dark matter present to hold them together. Since, almost 
all the evidences suggest that clusters of galaxies are 
stable configuration. Hence it was concluded that the 
clusters consist of both luminous and non-luminous matter,  
which was termed as dark matter. 

In this paper we have discussed about the dark matter, 
various experimental hints and evidences for dark matter, 
its composition, the role of neutrinos in dark matter 
formation and understanding of its dynamics. 
Baryon to photon ratio shows that our 
Universe is not closed by baryonic matter, which gives a 
clear indication of the existence of dark matter.  
Given the properties, neutrinos fit to be a strong candidate constituent for  
the dark matter as they have an advantage over other dark matter candidates, e.g. they 
cluster on large scale where the dark matter is needed to 
hold the large clusters of galaxies. Despite the weakness of interactions and smallness of masses, they 
can play an important role in cosmology.

In addition to three active flavours of neutrinos, there 
could also exist extra massive neutrino states that are 
sterile, i.e. they are singlets of the Standard Model 
gauge group and thus insensitive to weak interactions. 
Most of the current data on neutrino oscillations can 
perfectly be 
explained with only three active species, but there 
exist a few experimental results \cite{l1}-\cite{l6} that cannot be explained 
in this framework. If neutrino oscillations are 
responsible for all the experimental data, a solution 
might require additional (sterile) neutrino species. These 
kind of particles are predicted by many theoretical models 
beyond the SM \cite{g1}. Their masses are usually heavy, while 
lighter sterile neutrinos are rarer but possible. 
Recent studies propose sterile 
neutrino with a mass of the order of a few keV's and a 
very small mixing with the active neutrinos. Such heavy 
neutrinos could be produced by active-sterile oscillations 
but not fully thermalized, so that they could play the 
role of dark matter and replace the usual CDM component. 
But due to their large thermal velocity  (smaller than 
that of active neutrinos), they would behave as WDM and 
erase small-scale cosmological structures. 
At present the neutrino physics and neutrino astrophysics 
and cosmology are at the 
cross roads. On the one hand, it is impossible to deny 
that neutrinos oscillate and thus presumably have small 
masses, and on the other unless a sterile neutrino truly 
exists, there is a sense that neutrino masses are too small 
to be of very much cosmological interests.

In the galaxy formation scenario, galaxies can only form 
by the collapse of super-clusters. The detailed study shows 
that the collapse of super-clusters only happens very late 
and may be in contradiction with the existence of quasars 
of large red shift. Although the evidences for dark matter 
is wide and deep and existence of dark matter is based on 
the assumption that the laws of motion and gravity as 
formulated by Newton and extended by Einstein apply. On 
the other hand the modification in the theory of gravity 
can explain the effects attributed to dark matter and some 
scientists have proposed MOND (Modified Newtonian 
Dynamics). According to this theory at very low 
acceleration, corresponding to large distances, the usual 
law of gravitation is modified. Although MOND has had some 
success in explaining observations of galaxies, but failed 
to explain the observation of Bullet Clusters. So we need 
more experimental evidences to give a conclusive theory of 
dark matter.

\section*{Acknowledgments}
We thank Debasish Majumdar, SINP, for giving useful inputs. The Inter-University Centre for Astronomy \& Astrophysics (IUCAA), Pune is also acknowledged for providing research facilities during the completion of this work.

\end{document}